\documentclass[twocolumn,floatfix,superscriptaddress,amsmath,showpacs,showkeys,aps,prb]{revtex4}
\usepackage[utf8]{inputenc}
\usepackage[final]{graphicx}
\usepackage{t1enc}
\usepackage{bm}

\begin{document}

\bibliographystyle{apsrev}

\title{Domain walls, canted states and stripe width variation  in ultrathin magnetic films with perpendicular anisotropy}

\author{Santiago A. Pigh\'{\i}n}
\email{pighin@famaf.unc.edu.ar}
\affiliation{Instituto de Física de la Facultad de Matemática, Astronomía y Física (IFFAMAF-CONICET),
Universidad Nacional de Córdoba\\
Ciudad Universitaria, 5000 Córdoba, Argentina}

\author{Orlando V. Billoni}
\email{billoni@famaf.unc.edu.ar}
\affiliation{Instituto de Física de la Facultad de Matemática, Astronomía y Física (IFFAMAF-CONICET),
Universidad Nacional de Córdoba\\
Ciudad Universitaria, 5000 Córdoba, Argentina}
\author{Daniel A. Stariolo}
\email{stariolo@if.ufrgs.br}
\affiliation{Departamento de Física,
Universidade Federal do Rio Grande do Sul
and
National Institute of Science and Technology for Complex Systems\\
CP 15051, 91501-970 Porto Alegre, RS, Brazil}
\altaffiliation{Research Associate of the Abdus Salam International Centre for
Theoretical Physics, Trieste, Italy}
\author{Sergio A. Cannas}
\email{cannas@famaf.unc.edu.ar}
\affiliation{Instituto de Física de la Facultad de Matemática, Astronomía y Física (IFFAMAF-CONICET),
Universidad Nacional de Córdoba \\
Ciudad Universitaria, 5000 Córdoba, Argentina}

\date{\today}

\begin{abstract}
Stripe width variation in ultrathin magnetic films is a well known
phenomenon still not well understood. We analyze this problem
considering a 2D Heisenberg model with ferromagnetic exchange
interactions, dipolar interactions and perpendicular anisotropy,
relevant e.g. in Fe/Cu(001) films.  By extending a classic result of
Yafet \& Gyorgy (YG) and using Monte Carlo simulations we calculate the complete zero temperature phase diagram of the model. Through this calculation we analyze the correlation between domain walls structure and stripe width variation, as the perpendicular anisotropy changes. In particular, we found evidences that the recently detected canted state becomes the ground state of the system   close to the Spin Reorientation Transition (SRT) for any value of the exchange to dipolar couplings ratio.  Far away of the SRT the canted ground state is replaced by a saturated stripes state, in which in--plane magnetization components are only present inside the walls. We find that the domain wall structure strongly depends on the perpendicular anisotropy: close to SRT it is well described by YG approximation, but a strong departure is observed in the large anisotropy limit. Moreover, we show that stripe width variation is directly related to domain wall width variation with the anisotropy.
\end{abstract}

\maketitle
\section{Introduction}

Pattern formation in ferromagnetic
thin films with perpendicular anisotropy   and its  thermodynamical description  have been the subject of
intense experimental~\cite{AlStBi1990,
VaStMaPiPoPe2000,PoVaPe2003,WuWoSc2004,WoWuCh2005,ChWuWoWuScDoOwQi2007,PoVaPe2006,Portmann2006,ViSaPoPePo2008,CoCeOlTiViBa2008,AbVe2009}, theoretical~\cite{GaDo1982,YaGy1988,
CzVi1989,PePo1990,KaPo1993,AbKaPoSa1995,Po1998,StSi2001,BaSt2007} and numerical~\cite{DeMaWh2000,CaStTa2004,CaMiStTa2006,
RaReTa2006,GiLeLi2006,PiCa2007,NiSt2007,CaMiStTa2008,CaBiPiCaStTa2008,WhMaDe2008} work in the last 20 years.  Magnetic order in ultrathin ferromagnetic films is very complex due to the competition between exchange and dipolar interactions on different
length scales, together with a strong influence of shape and
magnetocrystalline anisotropies of the sample. These in turn are very susceptible to the  growth conditions of the films~\cite{Portmann2006,VaBlLa2008}.

Among the different magnetization patterns that have been observed in
these systems,  striped order (i.e., modulated patterns of local
perpendicular magnetization with a well defined half-wavelength or
stripe width $h$) at low temperatures is an ubiquitous phenomenon. One
intriguing fact  is the strong variation displayed by the equilibrium
stripe width $h$ in many of these systems, when either the temperature
or the film thickness is
changed\cite{PoVaPe2003,PoVaPe2006,WuWoSc2004,WoWuCh2005,ViSaPoPePo2008,CoCeOlTiViBa2008}.
The origin of such variation is still controversial, but recent
results suggest that a key point to understand it is the role played
by  the interfaces (i.e., the domain walls) between
stripes\cite{ViSaPoPePo2008}. Thus, a starting point to study this
problem is to compare the energies of striped patterns with different
domain wall configurations. An accurate description of the domain
walls requires  to take into account not only the perpendicular
component of the local magnetization, but also the in--plane
component. Indeed, some experimental results\cite{WuWoSc2004} are
consistent with the presence of Bloch domain walls, as expected for
perpendicularly oriented magnetization domains\cite{Be1998}.

 To
compute the energy contribution of domain walls it is important to
consider explicitly the out of plane anisotropy, together with the
exchange and dipolar interactions, whose competition is the
responsible for the appearance of striped patterns. A minimum model
that contains all these ingredients  is the 2D dimensionless
Heisenberg Hamiltonian:
\begin{widetext}
\begin{equation}
{\cal H} = -\delta \sum_{<i,j>} \vec{S}_i \cdot \vec{S}_j +
\sum_{(i,j)} \left[ \frac{\vec{S}_i \cdot \vec{S}_j }{r_{ij}^3} - 3 \,
\frac{(\vec{S}_i \cdot \vec{r}_{ij}) \; (\vec{S}_j \cdot \vec{r}_{ij})}{r_{ij}^5} \right]
- \eta \sum_{i} (S_i^z)^2
\label{hamiltoniano}
\end{equation}
\end{widetext}
where $\vec{S}_i$ are classical unit vectors, the exchange and anisotropy constants are normalized relative to the dipolar coupling
constant ($\delta \equiv J/\Omega, \eta \equiv K/\Omega$), $<i,j>$
stands for a sum over nearest neighbors pairs of sites in a  square
lattice, $(i,j)$ stands for a sum over {\it all distinct} pairs and
$r_{ij}\equiv |\vec{r}_i - \vec{r}_j|$ is the distance between spins
$i$ and $j$. 

In the large anisotropy limit $\eta\to\infty$ this model
reduces to an Ising model with short range ferromagnetic  and long
range antiferromagnetic interactions, whose ground state is the
striped one\cite{GiLeLi2006}. In that limit the  stripe width
increases exponentially with the exchange to dipolar coupling
ratio\cite{DeMaWh2000,PiCa2007} $h\sim\exp (\delta/2)$. For low values
of the anisotropy, the ground state of this model changes to a planar
ferromagnetic state\cite{YaGy1988}. In a classic work, Yafet and
Gyorgy computed the energy of striped domain configurations with Bloch
domain walls, by considering a sinusoidal (perpendicular)
magnetization profile at the walls and saturated magnetization inside
the stripes\cite{YaGy1988}. They found that above certain threshold
value $\eta>\eta_{min}$ a striped configuration has less energy than a
uniformly (in--plane) magnetized one. At this point the system shows a
Spin Reorientation Transition (SRT). Also within this approximation
the stripe width shows an exponential increase with the anisotropy
strength, while the domain wall width decreases algebraically for
large enough values of $\eta$, at least for large values of
$\delta$. This approximation is expected to work well close to the
SRT, where the effective anisotropy is small\cite{Po1998}. However,
for large values of $\delta$ and $\eta$ not too close to
the SRT  (i.e., when the width of both stripes and walls are large)
one can expect a large departure in the wall energy contribution
with respect to the true domain wall configuration, which should  approach a hyperbolic tangent profile\cite{Be1998}.

Another important point concerns the in--plane magnetization component
inside the stripes.
In their work Yafet and Gyorgy considered striped solutions where the
only in--plane components lay inside the walls, although they pointed
out how to extend their calculations to consider non--saturated
magnetization inside the stripes\cite{YaGy1988}. Based on that
calculation, Politi\cite{Po1998} reported that, at least for large
enough values of $\delta$, the magnetization should show an abrupt
saturation very close to the SRT, suggesting that the in--plane
component inside the stripes is not relevant. However, recently
Whitehead et al\cite{WhMaDe2008} obtained evidences of a
non--saturated ground state at a relatively small value of $\delta$
for a wide range of values of the anisotropy strength $\eta$. Their
numerical results also showed  a correlation between the stripe width
and the in--plane component variations and suggest that this ground
state configuration stabilizes at finite temperature, giving rise to
what they called a ``canted" phase. Hence, it is important to revise
the  zero temperature phase diagram of this model in the whole
$(\delta,\eta)$ space, including canted  configurations, in order to determine to what extent they can be relevant to real systems and their possible influence to the stripe width variation phenomenon.

It is worth noting that the perpendicular anisotropy changes
(inversely) with the film thickness\cite{CaBiPiCaStTa2008}. It has  also 
been pointed out that the changes in the film thickness can  act
as a change in the effective temperature\cite{PoVaPe2003}. Hence,
understanding the variation of the equilibrium stripe width and the
associated domain wall structure as the anisotropy changes at zero
temperature  can be of great help to understand the corresponding
properties a finite temperature, a far more complex problem.

In this work we analyze the complete equilibrium phase diagram at zero
temperature in the $(\delta,\eta)$ space of Hamiltonian
(\ref{hamiltoniano}). Following Yafet and Gyorgy's work, we  only
consider straight domains, i.e., domains in which the spin orientation
can be modulated along the $x$ direction but is constant in the
perpendicular direction $y$ ($(x,y)$ are the coordinates on the
plane of the film ). We  also consider only Bloch walls, i.e., walls
in which the magnetization stays inside the $yz$ plane. In order to
clarify notation and units, and to include some extensions,   we
briefly review Yafet \& Gyorgy's approximation  in Appendix
\ref{yafet}. To obtain the phase diagram we  compute the energy of
different types of magnetization profiles and compare them with
simulation results obtained through a zero temperature Monte Carlo
(MC) method specially designed for the present purposes. The MC method
is presented in Appendix \ref{MCmethod}. In section
\ref{energysection} we derive a general expression for the energy of striped magnetization profiles. The zero temperature phase diagram and associated properties are derived in section \ref{phasediagram}. A discussion and conclusions are presented in section \ref{discu}.

\section{Energy of striped magnetization profiles}
\label{energysection}

Let us consider a square lattice with $N=L\times L$ sites,
characterized by the integer indexes $(x,y)$, where $-L/2 \leq x \leq
L/2$ and $-L/2 \leq y \leq L/2$, in the limit $L\to\infty$. Hence, the
index $i$ in Eq.(\ref{hamiltoniano}) denotes a pair of coordinates
$(x,y)$. Following YG approximation\cite{YaGy1988}, we  consider only
uniformly magnetized solutions along every vertical line of sites,
i.e. $\vec{S}_{(x,y)}=\vec{M}(x)$, $\forall y$ and allow only Bloch
walls between domains of perpendicular magnetization,
 i.e. $M^x(x)=0$ $\forall x$. Therefore,

\begin{equation}\label{module}
    \left|\vec{M}(x) \right|^2 = \left[M^z(x) \right]^2+ \left[M^y(x) \right]^2=1.
\end{equation}

Then, for every value of $x$ there is only one independent component of the magnetization.

YG showed that for these types of  spin configurations the energy per spin can be mapped onto the  energy of a one dimensional XY model. The energy difference between an arbitrary magnetization profile $\vec{M}(x)$ and a uniformly in--plane magnetized state is then given by (see Appendix \ref{yafet-energy}):

\begin{widetext}
\begin{equation}
e\left[\vec{M}(x)\right]= (\delta-2c_2) - \frac{\delta'}{L}
\sum_x \vec{M}(x). \vec{M}(x+1) + \frac{1}{L} \sum_{x,x'}
\frac{M^z(x)\, M^z(x')}{|x-x'|^2}  - \frac{\kappa'}{L} \sum_x
\left[M^z(x) \right]^2 + C
\label{energia1}
\end{equation}
\end{widetext}

\noindent where $\delta'=\delta-2\, c_1$, $\kappa'=\eta-3\, g$, $c_1=0.01243\ldots$, $c_2=0.07276\ldots$, $g=1.202057\ldots$ and

\begin{equation}\label{correction}
    C \equiv C\left[M^y(x)\right]= 2(c_2-c_1)\frac{1}{L} \sum_x M^y(x)\, M^y(x+1)
\end{equation}

\noindent Although small, this correction term makes a non negligible
contribution when the domain walls are of the same order of the
lattice constant. This happens for small values of $\delta$ ($\delta <
5$), where both the stripe and wall widths are of the order of a few
lattice spacings. For larger values of $\delta$ it is reasonable to assume a smooth magnetization profile\cite{YaGy1988} $M^y(x+1)\approx M^y(x)$, so that

\begin{equation}\label{correction-smooth}
    C \approx 2(c_2-c_1)\frac{1}{L} \sum_x \left[ 1-  \left(M^z(x)\right)^2\right]
\end{equation}

\noindent can be absorbed into the anisotropy term in
Eq.(\ref{energia1}),  replacing $\kappa'\to\kappa= \eta-3\, g+2(c_2-c_1)$.

Considering now a stripe-like periodic structure of the magnetization profile  with period $2h$, i.e.

\begin{eqnarray}
 M^z(x+h)&=& -M^z(x), \label{antiperiodic}
 \end{eqnarray}

\noindent we can make use of a Fourier expansion:

\begin{equation}\label{Fourier-stripes}
    M^z(x) = M_0 \sum_{m=1,3,\ldots} b_m \; \cos{ \left(\frac{m\pi\, x}{h} \right)},
\end{equation}

\noindent where we have assumed $M^z(x)$ an even function of $x$ just for simplicity.

 The anisotropy term in Eq.(\ref{energia1}) can be written as

\begin{equation}\label{ean}
    e_{an}=- \frac{\kappa'}{2h} \sum_{x=1}^{2h} \left[M^z(x) \right]^2= - \frac{\kappa'\, M_0^2}{2} \sum_{m=1,3,\ldots} b_m^2,
\end{equation}

\noindent and the dipolar term\cite{YaGy1988}

\begin{equation}
    e_{dip}=\frac{1}{L} \sum_{x,x'}
\frac{M^z(x)\, M^z(x')}{|x-x'|^2}=M_0^2 \sum_{m=1,3,\ldots} b_m^2  D_m(h),
    \label{eDipInf}
\end{equation}

\noindent where\cite{GrRi1994}

\begin{equation}\label{Dm}
    D_m(h) \equiv \sum_{u=1}^\infty \frac{\cos{(m\pi\, u/h)}}{u^2}= \frac{\pi^2}{6} - \frac{\pi^2m}{2h} + \left( \frac{\pi m}{2h}\right)^2.
\end{equation}

In the general case, it is better to let the exchange term (and the correction term $C$ as well) in Eq.(\ref{energia1}) expressed in terms of the angle $\phi(x)$ between $\vec{M}(x)$ and the $z$ axis:

\begin{eqnarray}
  M^z(x) &=&  \cos{[\phi(x)]}\\
  M^y(x) &=& \sin{[\phi(x)]}
\end{eqnarray}

\noindent where the angle $\phi(x)$ has the same periodicity of $\vec{M}(x)$.
We have that

\begin{equation}
    e_{exc}=-\delta \frac{1}{L} \sum_x \cos{\left[ \phi(x)-\phi(x+1)\right]}.
\end{equation}

Putting all the terms together we get the general expression:

\begin{widetext}
\begin{equation}
e\left[\vec{M};\delta,\eta \right]= (\delta-2c_2) - \delta'
\frac{1}{L} \sum_x \cos{\left[ \phi(x)-\phi(x+1)\right]} + M_0^2
\sum_{m=1,3,\ldots} b_m^2  D_m(h) - \frac{\kappa'\, M_0^2}{2}
\sum_{m=1,3,\ldots} b_m^2 + C .
\label{energiabm}
\end{equation}
\end{widetext}

\section{Zero temperature phase diagram}
\label{phasediagram}

In this section we  look for the minimum of Eq.(\ref{energiabm}) for
different values of $\delta,\eta$. We propose  different  striped
magnetization profiles $M^z(x)$ and compare the energies obtained by
minimizing Eq.(\ref{energiabm}) for each profile with respect to variational parameters.

\subsection{Small values of $\delta$: Sinusoidal Wall magnetization Profile (SWP) approximation}
We first consider a profile as proposed by YG, that is constant inside
the stripes and has a sinusoidal variation inside the walls between
stripes (see Fig.1 in Ref.~\onlinecite{YaGy1988}):

\begin{widetext}
\begin{equation}
    M^z(x)=\left\{\begin{array}{ll}
             M_0 & \mbox{if} \;\; 0 \leq x \leq \frac{h-w}{2} \\
             M_0\, \cos{\left( \frac{\pi (x-(h-w)/2)}{w}\right)} & \mbox{if} \;\; \frac{h-w}{2} \leq x \leq \frac{h+w}{2} \\
             - M_0 & \mbox{if} \;\; \frac{h+w}{2} \leq x \leq h
           \end{array} \right.\label{perfilYG}
\end{equation}
\end{widetext}

\noindent where $M_0$ is the absolute value of magnetization inside
the stripes and $w$ is the wall width.  In order to allow for canted
profiles, we take $M_0= \cos \theta$, where $\theta$ is the canted
angle, i.e. we define it as the minimum angle of the local
magnetization with respect to the $z$ axis. Yafet and Gyorgy solved
this variational problem for $M_0=1$ in the continuum
limit\cite{YaGy1988}, i.e. when $h\gg 1$ and $w\gg1$, so that the
profile can be considered a smooth function of $x$. While this
approximation is expected to work well for large enough values of
$\delta$, it breaks down for relatively small values of it, where the
discrete character of the lattice has to be taken into
account. However, the variational problem for that range of values of
$\delta$ can be solved exactly (although numerically) by minimizing
Eq.(\ref{energiabm}) with respect to the {\it integer} variational
parameters $h$ and $w$ and continuous parameter $\theta$. In other
words, for every pair of values $(\delta,\eta)$ we evaluate the energy
Eq.(\ref{energiabm}) for the profile (\ref{perfilYG}) with different
combinations of $h=1,2,\ldots$ and $w=1,2,\ldots$ within a limited
set. For every pair of values $h,w$,we look for the value of $\theta$
that minimizes the energy with a resolution $\Delta \theta = 0.01$ and
compare all those energies. Details of that evaluation are given in
Appendix \ref{YGdiscrete}.
This  calculation is feasible for values  up to $\delta=10$, for which the maximum value of $h$ (bounded by the stripe width in the $\eta\to \infty$ limit) remains relatively small (smaller than $h=140$). Some results for $\delta=12$ close to the SRT were also obtained. All the results of this calculation are compared against Monte Carlo (MC)  simulations. Details of the MC method used are given in Appendix \ref{MCmethod}. Through these calculations we obtain a zero temperature phase diagram for low values of $\delta$.

Before presenting the results, it is important to introduce some
notations and definitions of different types of solutions. We
distinguish between four types of solutions. If the minimum energy
solution corresponds to $w=1$ and $\theta=0$ (within the resolution
$\Delta \theta$), we call this a {\it Striped Ising Profile} (SIP),
i.e. a square wave like profile. If $\theta=0$ but $w > 1$, we call
this a {\it Saturated State}. These states show a finite parallel
component of the magnetization inside the walls. If $ 0 < \theta < \pi/2$ the solution is
a {\it Canted State}. Finally, if $\theta = \pi/2$ ($M_0=0$) we have a {\it Planar Ferromagnet} (PF).

\begin{figure}
\includegraphics[scale=0.3]{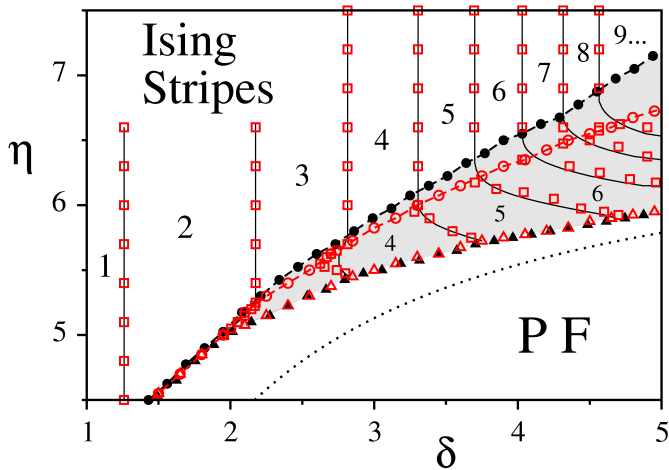}
\caption{\label{pd1} (Color online) Zero temperature phase diagram for
  small values of $\delta$. Black filled symbols and black solid
  lines: MC simulations. Open red symbols: SWP approximation. Squares
  and continuous black lines correspond to transition lines between
  striped states of different width. The shaded region corresponds to
  the Canted State ($0 < \theta < \pi/2$).
Triangles are transition lines between  Planar Ferromagnet
  ($\theta=\pi/2$) and Canted States (Spin Reorientation Transition
  line). Circles  mark transitions between the Canted and the Stripes
  Ising state ($\theta=0$ and $w=1$). Notice the excellent agreement
  between the MC and SWP calculations close to the SRT, while the SWP
  approach underestimates the transition line between the canted and
  Ising Stripes states. The dotted line corresponds to the contimuum
  approximation of YG for the SRT (Eq.(\ref{etaSRT})).}
\end{figure}

The zero temperature phase diagram for small values of $\delta$
($\delta \leq 5$) is shown in Fig.\ref{pd1}. For relatively large
values of $\eta$ the minimum energy configuration is always the Ising
one (SIP), with a stripe width independent of $\eta$. For small values
of $\eta$ the minimum energy configuration is the PF, with a spin
reorientation transition line (SRT), either to the Ising state for $h
<3$ ($\delta\sim 2)$) or to a canted one for $h\geq 3$ ($\delta >
2$). No Saturated State configurations are observed for $\delta<6$.

Inside the canted region, a strong stripe width variation with the
anisotropy is observed at constant $\delta$ . Note that the vertical
lines that separate Ising striped states with consecutive values of
$h$ bend inside the canted region and become almost horizontal as $\delta$
increases. Hence, the exponential increase of $h$ with $\delta$ in the
Ising region (vertical lines) changes to an exponential increase with
$\eta$ deep inside the canted region (curved lines on the right of Fig.\ref{pd1}).

\begin{figure}
\includegraphics[scale=0.3]{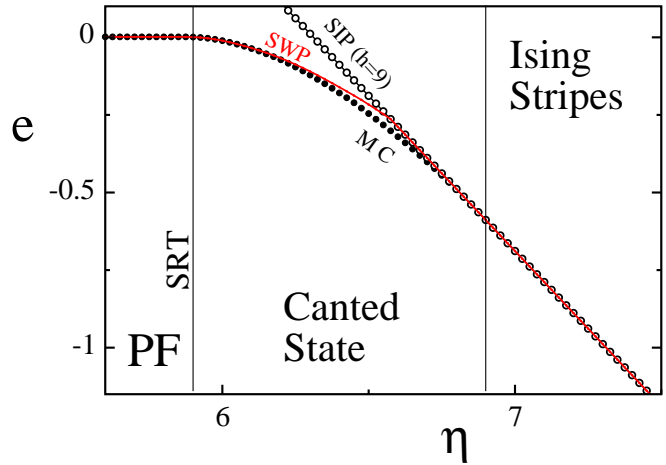}
\caption{\label{energyYGvsMC} (Color online) Energy per spin (with respect to the PF state) vs. $\eta$ for $\delta=4.58$ within the SWP (full red line) and MC calculations (filled black circles). Open black circles correspond to the energy of a Striped Ising Profile (SIP) with $h=9$ (equilibrium value for $\eta\to\infty$).}
\end{figure}

We also find an excellent agreement between the sinusoidal wall
profile approximation  and the MC results, except close to the
transition between the Ising and the canted states. Such disagreement
is due to the fact that the actual wall is not well described by a
sinusoidal profile far away of the SRT line, as will be shown later. In Fig.\ref{energyYGvsMC} we show a comparison between the energy of the SWP and the MC results  as a function of $\eta$ for $\delta=4.58$. The range of values of $\eta$ where the  walls are not well described by a sinusoidal profile increases with $\delta$.

For large enough values of $\delta$ the variational problem for the SWP can be solved in a continuum approximation introduced by YG\cite{YaGy1988}. This leads to the equations (see Appendix \ref{YGcontinuum}):

\begin{eqnarray}
   \frac{\delta\, k}{\Delta} &=& \frac{\pi}{2}\, G(\Delta) (1+\sin{\theta}) \label{YGEq1-2}\\
  \frac{\delta\, k^2}{\Delta^2} &=& - \left[ \gamma + \pi\, k\, \frac{dG}{d\Delta}\right] (1+\sin{\theta}) \label{YGEq2-2}\\
  \frac{\delta\, k^2}{2\Delta} &=&  - \left[ 2\gamma
    \left(1-\frac{\Delta}{2} \right)- \pi\, k\, G(\Delta)\right]
  \sin{\theta}, \label{YGEq3'}
\end{eqnarray}

\noindent where $\Delta\equiv w/h$, $k\equiv \pi/h$ and $\gamma=\pi^2/3-\kappa$.  In the limit $\Delta\to 1$ (pure sinusiodal profile) these equations can be solved analytically and  predict a SRT at the line

\begin{equation}\label{etaSRT}
    \eta_{SRT}(\delta)= a- \frac{\pi^2}{2\delta}
\end{equation}

\noindent with $a=\pi^2/3+3g-2(c_2-c_1)$ (see Appendix \ref{YGcontinuum}). The line Eq.(\ref{etaSRT}) is also depicted in Fig.\ref{pd1}. Notice the disagreement between the continuum approximation and the exact one for $\delta \leq 5$. This discrepancy becomes smaller than $1\%$ only for $\delta > 7$.

\begin{figure}
\includegraphics[scale=0.42]{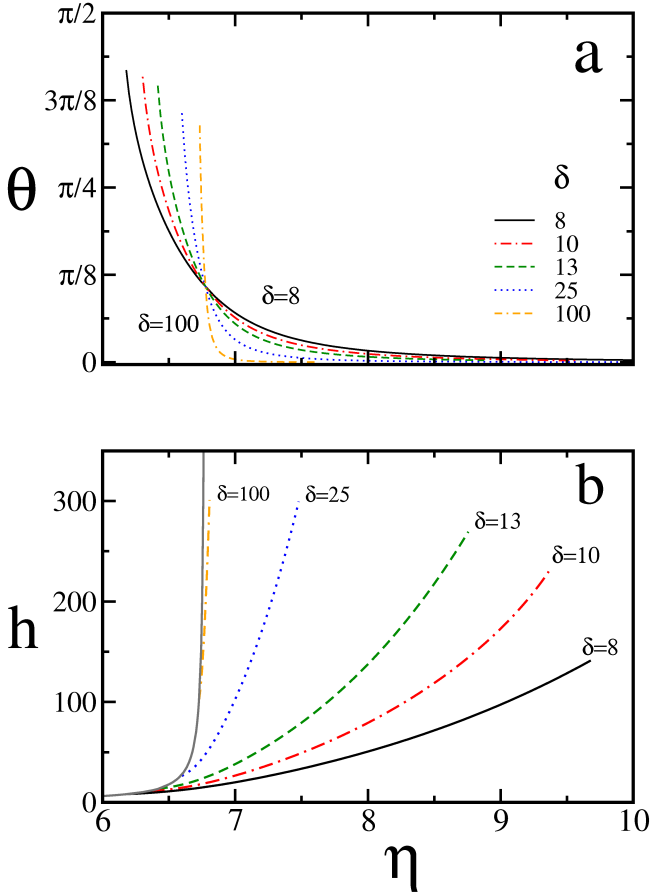}
\caption{\label{YG-cont-num} (Color online) (a) Canted angle $\theta$
  and (b) equilibrium stripe widht $h$, as a function of the
  anisotropy $\eta$ for different values of $\delta$ from the
  continuum approximation of the SWP
  Eqs.(\ref{YGEq1-2})-(\ref{YGEq3'}). The full grey line at the left
  corresponds to the SRT borderline given by Eq.(\ref{etaSRT}) with
  $h=\delta$ (see Appendix \ref{YGcontinuum}).}
\end{figure}

For arbitrary values of $\eta$ and $\delta$ Eqs.(\ref{YGEq1-2})-(\ref{YGEq3'})  can be solved numerically. In Fig.\ref{YG-cont-num} we show the numerical solutions for $\theta$ and $h$ as a function of $\eta$ for different values of $\delta$. We see that the range of values of the anisotropy $\eta$ for which the canted angle is appreciable different from zero within this approximation is strongly depressed as $\delta$ increases. For values $\delta\sim 100$ the canted configuration almost disappears, except very close to the reorientation line, consistently with the results reported by Politi\cite{Po1998}.

\begin{figure}
    \includegraphics[scale=0.28,angle=0]{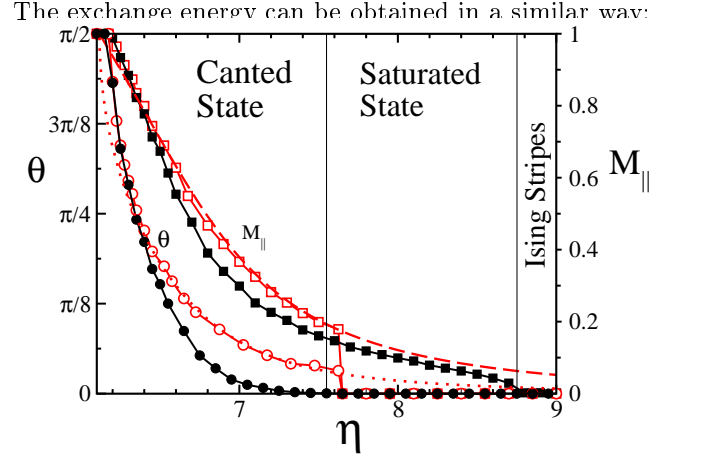}
    \caption{\label{mxSvsEta} (Color online) Canted angle (circles) and in-plane
      magnetization (squares) as a function of $\eta$ for
      $\delta=7.5$. Filled black symbols correspond to MC
      calculation. Open red symbols  corresponds to the discrete SWP
      approximation, while the red doted and dashed lines correspond
      to the continuum (YG) approximation of the SWP. Continuous black and red lines are only a guide to the eye.}
  \end{figure}

Indeed, from our MC simulations, we observe that the range of values
for which the canted state has the minimum energy gradually shrinks as
$\delta$ increases, being replaced by a saturated state for values of
$\eta$ above certain threshold. This can be observed in
Fig.\ref{mxSvsEta}, where we show the behavior of the canted angle and
the in--plane magnetization component $M_{||}=(1/L)\sum_x M^y(x)$ as a
function of $\eta$ for $\delta=7.5$. The Monte Carlo data shows the
existence of a wide range of values of $\eta$ for which the canted
angle is zero while $M_{||}\neq 0$, meaning that the non null
in--plane components are concentrated inside the walls. In other
words, in that region we have a saturated state with thick walls $w
>1$. Notice also that the SWP approach completely fails to describe
those states. Moreover, we observe from our MC simulations that the
SWP cease to be the minimum energy solution for values of $\eta$
relatively close to the SRT, well before the saturated state sets up (see Fig.\ref{mxSvsEta}). This effect becomes more marked as $\delta$ increases.

\subsection{Large values of $\delta$: Hyperbolic Wall magnetization Profile (HWP) approximation}

\begin{center}
  \begin{figure}
    \includegraphics[scale=0.3,angle=0]{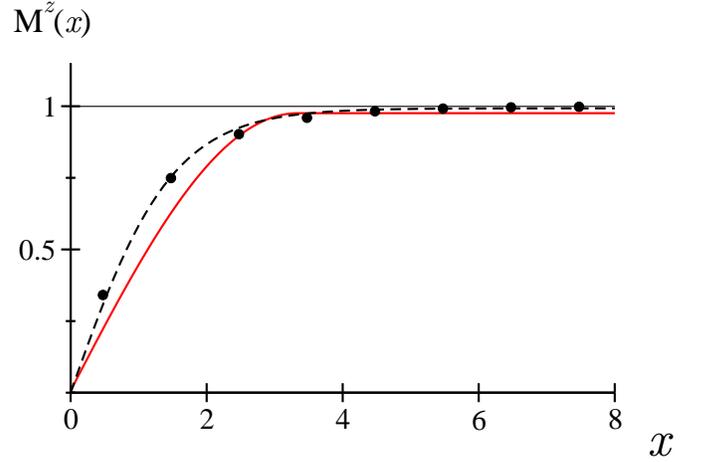}
    \caption{\label{profiletanh} (Color online) Magnetization profile
    in a saturated state ($\theta=\pi/2$ and $w>1$) for $\delta=8$ and
    $\eta = 7$. Black symbols correspond to the MC simulations
    ($M_0=0.999$ and $h=20$). The full red line corresponds to the YG
    approximation for the SWP ($M_0=0.98$, $w=6.7$ and $h=20$). The
    black dashed line is a fit using a hyperbolic tangent function $M_0\, \tanh (x/l_w)$ ($M_0=0.993$ and  $l_w=1.48$).}
  \end{figure}
\end{center}

 As already pointed out, the actual magnetization profile departs from
 the SWP for large values of $\eta$ and $\delta$. This is expected
 from micromagnetic theory, which  in that limit predicts that the
 wall structure will be dominated by the interplay between anisotropy and exchange, leading to an hyperbolic tangent shape of the wall\cite{Be1998}. This can be observed in Fig.\ref{profiletanh}. Hence we considered a periodic magnetization profile with hyperbolic tangent walls (HWP) defined, for a wall centered at $x=0$, by

\begin{equation}\label{hyperbolicprofile}
    M^z(x)= M_0\, \tanh{\left(\frac{x}{l_w} \right)} \;\;\;\; \mbox{for} \;\;\;\; -h/2 \leq x\leq h/2,
\end{equation}

\noindent together with Eq.(\ref{antiperiodic}) where $M_0=\cos{\theta}$ as before. In the large $\delta$ limit, assuming a smooth profile  $h \gg 1$ and $l_w\gg 1$, the anisotropy energy can be expressed as:

\begin{eqnarray}
   e_{an}&=&- \frac{\kappa}{L} \sum_x
\left[M^z(x) \right]^2 \nonumber\\
&\approx& - \kappa\,M_0^2 \left[ 1-\frac{2\,l_w}{h}\tanh{\left( \frac{h}{2\, l_w} \right)}\right]. \label{anitanh}
\end{eqnarray}

\noindent The exchange energy can be obtained in a similar way:

\begin{eqnarray}
    e_{exc} &\approx& -\delta + \frac{\delta}{L} \sum_{x=1}^L \left( \frac{d\phi(x)}{dx}\right)^2\nonumber \\
 &\approx&-\delta +\frac{\delta}{h l_w} \int_{0}^{h/2l_w}
 \frac{\mbox{sech}^4u}{M_0^{-2} - \tanh{u}}du .
    \label{eexchHWP}
\end{eqnarray}

\noindent Solving the last integral we finally obtain

\begin{widetext}
\begin{equation}
    e_{exc}=-\delta \left[ 1 -  \frac{l_w}{h} \left( \frac{M_0^2-1}{M_0} \tanh^{-1}\left(M_0 \tanh \left( \frac{h}{2l_w} \right)\right) \\
+ \tanh \left( \frac{h}{2l_w} \right) \right) \right]. \label{eexchHWP2}
\end{equation}
\end{widetext}

\begin{figure}
\includegraphics[scale=0.32]{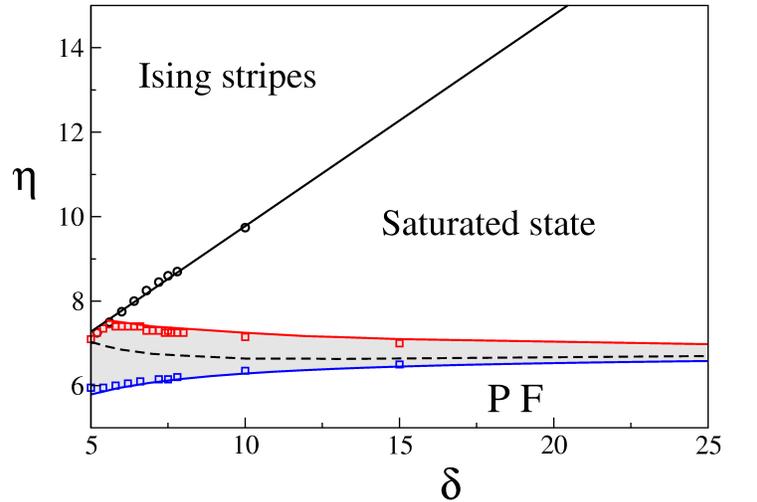}
\caption{\label{pd2} (Color online) Zero temperature phase diagram for
  large values of $\delta$. The shaded region corresponds to the Canted state. Symbols correspond to MC simulations and lines to theoretical results. The dashed line correspond to the crossover between sinusoidal and hyperbolic wall structure. The lower line (blue) corresponds to Eq.(\ref{etaSRT}). The middle line (red) is obtained from the HWP minimum energy solution with $\theta = 0.01$. The upper line (black) corresponds to Eq.(\ref{etatanhYG}).}
\end{figure}

The dipolar energy can be calculated using Eqs.(\ref{eDipInf}) and
(\ref{Dm}). The Fourier coefficients for the profile
(\ref{hyperbolicprofile}) can be computed using the approximation
(\ref{perfilTanh}) (see Appendix \ref{Fouriertangh}). This leads to an
expression for the total energy as a function of the variational
parameters $h$, $\theta$ and $l_w$ that can be minimized
numerically. Comparing the minimum energy solution for the SWP and the
HWP we obtain the crossover line between sinusoidal and hyperbolic
wall structure shown in Fig.\ref{pd2} (dashed line). Above that line
the HWP has always less energy than the SWP. We also calculated the
transition line between the canted and the saturated states by setting
the condition $\theta = 0.01$, to be consistent with the criterium
used in the MC calculations. The results are shown in Fig.\ref{pd2}
together with the SRT line Eq.(\ref{etaSRT}), and compared with MC
calculations up to $\delta=15$. The excellent agreement with the MC
results gives support to the analytic approximations.

For large values of $\eta$ the exponential increase of $h$ makes it
cumbersome to apply the approximation of Appendix \ref{Fouriertangh}
for the calculation of the dipolar energy. Instead of that, we can use
the following heuristic argument to obtain a reasonable
approximation. The main error introduced by the SWP approach is in the
exchange and anisotropy contributions to the energy. Since the main contribution to  the dipolar energy is given by the interaction between domains, we can assume that the dipolar contribution of the wall is relatively independent of its shape. Hence, we can approximate it by Eq.(\ref{edipYG}). Furthermore, taking $w=f\, l_w$ ($f$ is a fitting parameter of order one to be fixed later) in
the limit $\Delta \ll 1$ ($l_w/h \ll1$), $G(\Delta)$ is very well approximated by\cite{WuWoSc2004}

\begin{equation}\label{GDelta-approx}
    G(\Delta) \approx  \frac{8}{\pi^2}\, ln\left(\frac{6 \pi}{5\Delta} \right)
\end{equation}

\noindent Assuming then

\begin{equation}\label{dipolarenergyHWP}
    e_{dip}= M_0^2 \left[  \pi^2 \left(\frac{1}{3}\,- \frac{f\, l_w}{6 h}\right) - \frac{4}{h}\, \ln \left( \frac{6 \pi h}{5 f\,l_w} \right) \right],
\end{equation}

\noindent we compare the energy obtained with the above equation with that obtained using the approximation of Appendix \ref{Fouriertangh} for different values of the system parameters. We verified that the error made by the approximation Eq.(\ref{dipolarenergyHWP}) taking $f=4$ is always smaller than $1\%$ for $h/l_w \geq 20$. We also observe that the best agreement with the MC results is obtained for $f=4$. Assuming then $M_0=1$,
the total energy per spin (relative to the parallel magnetized state)  for the HWP can then be approached by

\begin{eqnarray}
    e_{HWP}&=& \frac{\pi^2}{3} -\kappa +\frac{\delta/l_w+2\,l_w(\kappa -\pi^2/3)}{h} \nonumber\\
    & & - \frac{4}{h}\, \ln \left( \frac{3\, \pi h}{10\,l_w} \right). \label{energyHWP}
\end{eqnarray}

\noindent Minimizing Eq.(\ref{energyHWP}) with respect to the variational
parameters $h$ and $l_w$ (using Eq.(\ref{GDelta-approx})) leads to:

\begin{equation}\label{htanhYG}
    h =  \frac{10}{3 \pi}\, l_w \exp \left[\frac{\delta}{2 l_w}\right],
\end{equation}

\noindent with

\begin{equation}\label{lwtanhYG}
    l_w  =  \frac{ \delta }{2 + \sqrt{ 4 + 2 (\kappa  -  \pi^2/3)\delta }},
\end{equation}

\noindent in agreement with a  derivation made by Politi\cite{Po1998}.

With the previous calculation we can also estimate the transition line between the saturated and the Ising Striped state. In the large $h$ limit the energy for a SIP, i.e. for

\begin{equation}\label{phi-ising}
    \phi(x)=\left\{\begin{array}{cc}
                    0 & \mbox{if} \;\; 0\leq x\leq h/2 \\
                    \pi & \mbox{if} \;\; h/2< x\leq h
                  \end{array}
    \right.
\end{equation}

\noindent can be easily calculated from Eq.(\ref{energiabm}). The Fourier coefficients can be obtained as the $\Delta\to 0$ limit of Eq.(\ref{bm-YG}):

\begin{equation}\label{bm-ising}
    b_m=(-1)^{(m-1)/2} \frac{4}{\pi m}.
\end{equation}

\noindent Using  Eq.(\ref{Dm}) the dipolar energy is then given by

\begin{equation}\label{edip-ising}
    e_{dip}\sim\frac{\pi^2}{3} - \frac{8}{h} \sum_{m=1,3,\ldots}^{2h-1} \frac{1}{m} + \frac{4}{h}\sim \frac{\pi^2}{3} +4\frac{\psi(h)-\beta}{h},
\end{equation}

\noindent where $\beta\equiv \gamma_e +\ln\, 4-1$, $\gamma_e\approx0.577216$ is the Euler gamma constant and  $\psi(x)$ is the digamma function\cite{GrRi1994}. The energy per spin respect to the in--plane magnetized state is then given by

\begin{equation}
e_I =  -\kappa' + \frac{\pi^2}{3}+\frac{2\,\delta'-\,\beta}{h} -\frac{4\,\psi(h)}{h}
\label{energiaIsing}
\end{equation}

\noindent Minimizing Eq.(\ref{energiaIsing}) with respect to $h$ leads to the equation $\delta'/2-\beta= F(h)$, where $F(h)= \psi(h)-h\,\psi'(h)\sim \ln h -1$, thus recovering the known result $h\sim e^{\delta/2}$. Comparing the energies, we find that the HWP  has less energy than the Ising state for any value of $\eta$. Eq.(\ref{htanhYG}) shows that the stripe width variation in the Saturated state is determined by the change in the wall width as the anisotropy increases. Hence, $h$ will change until the wall width reaches the atomic limit, i.e. for $l_w=1$, where Eq.(\ref{htanhYG}) recovers the Ising behavior  $h\sim e^{\delta/2}$. Imposing the condition $l_w=1$ to Eq.(\ref{lwtanhYG}) we obtain the transition line between the Saturated and the Ising Stripes states:

\begin{equation}\label{etatanhYG}
    \eta  =\frac{1}{2}\, \delta -2 + \frac{\pi^2}{3} + 3g -2(c_2-c_1),
\end{equation}

\noindent which is also shown in Fig.\ref{pd2}, in complete agreement with the MC results.

\begin{figure}
\includegraphics[scale=0.32]{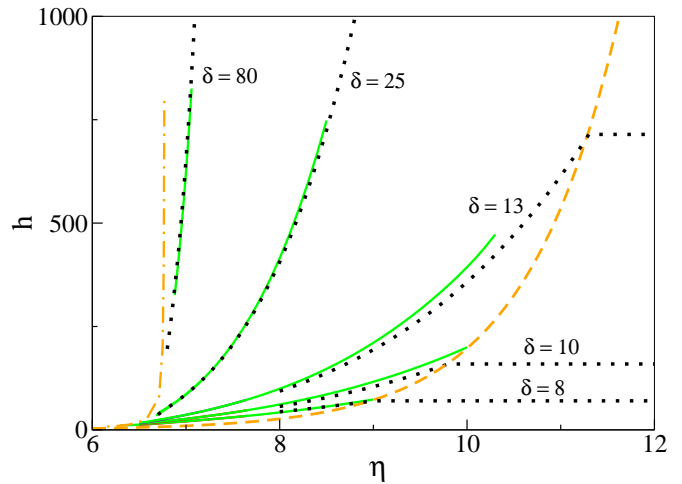}
\caption{\label{hvseta-HWP} (Color online) HWP equilibrium  stripe
  width $h$ vs. $\eta$ for different values of $\delta$. Full green
  lines correspond to the variational solution of Eqs.(\ref{anitanh})
  and (\ref{eexchHWP2}) using the approximation (\ref{perfilTanh}) for
  the Fourier coefficients in the dipolar energy. The dotted lines
  correspond to the asymptotic approximation given by
  Eqs.(\ref{htanhYG}) and (\ref{lwtanhYG}). The dash-dotted line
  corresponds to the SRT borderline given by Eq.(\ref{etaSRT}) with
  $h=\delta$. The dashed line corresponds to the borderline between
  Saturated and Ising Stripes states.}
\end{figure}

\begin{figure}
\includegraphics[scale=0.32]{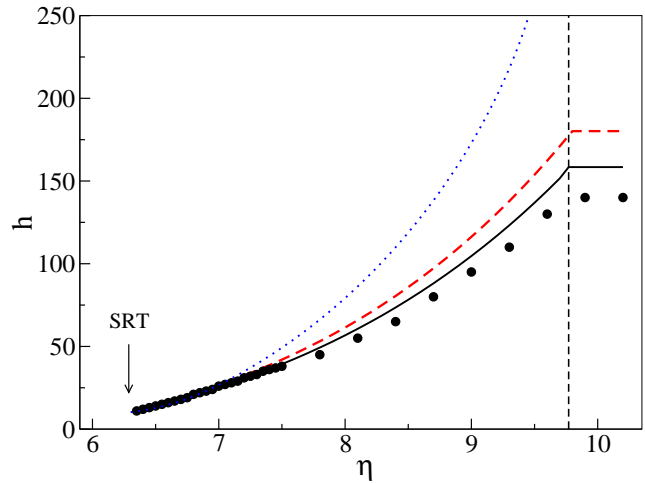}
\caption{\label{hvseta-delta10} (Color online) Comparison of the
  equilibrium  stripe width $h$ vs. $\eta$ obtained within the
  different methods for $\delta=10$. Symbols correspond to MC
  simulations. Full black  line corresponds to the asymptotic
  approximation for the HWP given by Eqs.(\ref{htanhYG}) and
  (\ref{lwtanhYG}). The red dashed line corresponds to the variational
  solution of Eqs.(\ref{anitanh}) and (\ref{eexchHWP2}) using the
  approximation (\ref{perfilTanh}) for the Fourier coefficients in the
  dipolar energy.  The blue dotted line corresponds to the continuous SWP. The vertical dashed line corresponds to the transition between Saturated and Ising Stripes states given by Eq.(\ref{etatanhYG}).}
\end{figure}

In Fig.\ref{hvseta-HWP} we show the variation of stripe width $h$
vs. $\eta$ for different values of $\delta$ in the HWP, comparing the
variational  solution from Eqs.(\ref{anitanh}) and (\ref{eexchHWP2})
using the approximation of Appendix \ref{Fouriertangh} and the
asymptotic approximation given by Eqs.(\ref{htanhYG}) and
(\ref{lwtanhYG}). In Fig.\ref{hvseta-delta10} we compare the
equilibrium  stripe width $h$ as a function of $\eta$ obtained within
the different approximations used in this work for $\delta=10$ and
with the MC simulations. Notice that the asymptotic approximation for
the HWP given by Eqs.(\ref{htanhYG}) and (\ref{lwtanhYG}) shows a
better agreement with the MC results than using the approximation (\ref{perfilTanh}) for the Fourier coefficients in the dipolar energy. This is because we adjusted the fitting parameter $f$ to optimize the agreement with the MC results at low values of $\delta$. From Fig.\ref{hvseta-HWP} we see that the discrepancy between both approximations becomes negligible in the large $\delta$ limit.

\section{Discussion and conclusions}
\label{discu}

The main results of this work are summarized in Figs.\ref{pd1} and
\ref{pd2}, which display the complete zero temperature phase diagram
of the model defined by the Hamiltonian (\ref{hamiltoniano}).
Working upon reasonable assumptions for the ground states, like
perfectly straight modulations in one dimension and Bloch domain
walls, we analyzed minimum energy configurations combining a
variational analysis with Monte Carlo results. We find four
qualitatively different kinds of solutions: a planar ferromagnet for
small anisotropies, and three types of perpendicular striped states: a
canted state where the local magnetization has a finite in-plane
component, a saturated state in which the in-plane component is
restricted to the domain walls, and an Ising stripe state with sharp
walls for large anisotropies.

The canted and staturated states give valuable information on the
behaviour of the stripe width as the anisotropy and exchange parameters
change, a still open and debated question\cite{WoWuCh2005,ViSaPoPePo2008}. We find that stripe width
variation is directly associated to the presence of finite width
domain walls. For large enough values of the anisotropy $\eta$ the
ground state of the system is always an Ising Striped state, no matter
the value of the exchange coupling $\delta$. In those states domain
walls are sharp and the stripe width is completely independent of
$\eta$. It grows exponentially with the exchange coupling.

At the SRT the system always passes through a canted state as the
anisotropy increases, although the range of values of $\eta$ where the
canted angle is different from zero narrows as $\delta$ increases. The
exchange to dipolar coupling  ratio in fcc Fe based ultrathin films
can be roughly estimated to be $\delta \sim 100$ (considering a cubic
bilayer of Fe/Cu(100), where\cite{WuWoSc2004} the exchange coupling
$J_{Fe}\sim 30\, meV$, the lattice constant $d_{Fe} \sim 2 ML$
and\cite{DuArMa1996} $\mu_{Fe}\sim 3\,\mu_B$). For $\delta \sim 100$
the anisotropy interval for the canted phase is approximately $\Delta \eta= \eta -\eta_{SRT} \approx 0.2$. Although narrow, this suggests that the canted phase should be detectable close enough of the SRT, in systems like low temperature grown\cite{EnPetReLiDmKe2003} Fe/Cu(100) or Fe/Ni/Cu films\cite{WoWuCh2005} (room temperature grown Fe/Cu(100) do not exhibit SRT\cite{PoVaPe2006}, suggesting a rather large value of the microscopic anisotropy).

  For $\delta < 6$ stripe width variation appears always together with a varying canted angle. Close enough to the SRT domain walls present a  sinusoidal shape in agreement with YG results, but as the anisotropy and the exchange increase, the wall profile changes to a hyperbolic tangent shape, as expected from micromagnetic calculations, while the magnetization inside the domains becomes fully saturated. For $\delta>6$ the ground state is given by the Saturated State, except very close to the SRT. A similar effect (i.e.  a crossover between a sinusoidal and a saturated magnetization profile) has been observed in room temperature grown fcc Fe/Cu(100) ultrathin films, as the temperature decreases from $T_c$, even though those systems do not present SRT\cite{ViSaPoPePo2008}.

 In the Saturated state, the stripe width increase with $\eta$  is
 directly related to the wall width decrease through the relation
 $h\sim e^{\delta/2l_w}$. The wall width in turns is determined by the
 competition between exchange and anisotropy. Once the anisotropy is
 large enough that the wall width reaches the atomic limit, $h$ growth
 stops. One may wonder whether a similar mechanism could be behind the stripe width variation with temperature.

Besides its direct application to real systems, knowing the ground state of this system for arbitrary values of the exchange coupling is of fundamental  importance to  have a correct interpretation of Monte Carlo simulation results. Being one of the more powerful tools to analyze these kind of systems at the present (specially at finite temperatures), it is basically limited by finite size restrictions, which implies relatively small values of $\delta$ (the characteristic length $h$ of the problem  grows exponentially with $\delta$ at low temperatures).

\section{Acknowledgments}

 We thank N. Saratz for advise about experimental results on  ultrathin magnetic films. This work was partially supported by grants from CONICET,
FONCyT grant PICT-2005 33305 , SeCyT-Universidad Nacional de C\'ordoba (Argentina),
CNPq and CAPES (Brazil), and ICTP grant NET-61 (Italy).

\appendix
\section{Yafet \& Gyorgy Approximation}
\label{yafet}

We briefly review in this appendix the derivation of the main results of YG approximation\cite{YaGy1988}.

\subsection{Energy per spin (Eq.(\ref{energia1}))}

\label{yafet-energy}

The expression for the exchange and anisotropy energies per spin  from Eq.(\ref{hamiltoniano}) in terms of the one dimensional magnetization profile $\vec{M}(x)$ is straightforward:

\begin{equation}\label{exc-ani}
    e_{ex}+e_{an}= -\delta - \frac{\delta}{L}
\sum_x \vec{M}(x). \vec{M}(x+1) - \frac{\eta}{L} \sum_x
\left[M^z(x) \right]^2
\end{equation}

The dipolar energy per spin can be expressed as $e_{dip}= e_{dip}^s+e_{dip}^{int}$, where $e_{dip}^s $ is the self-interaction energy (i.e., the sum over $x$ of the interaction energy between spins  belonging to the same line at $x$) and $e_{dip}^{int}$ is the  interaction energy between  all different pairs of lines. The self interaction term is given by\cite{YaGy1988}

\begin{equation}
e_{dip}^s  = -2\, g\, + \frac{3\, g}{L}\sum_x \left[M^z(x) \right]^2
\label{Es3}
\end{equation}

\noindent with

\begin{equation}\label{g}
    g= \sum_{n=1}^\infty \frac{1}{n^3}=\zeta(3)= 1.202057
\end{equation}

\noindent where $\zeta(x)$ is the Riemann Zeta function. The interaction term can be expressed as

\begin{equation}
e_{dip}^{int} = \frac{1}{L}\sum_{x\neq x'} E_{dip}^{int}(x,x')
\label{Es1}
\end{equation}

\noindent where the sum in the above expression is taken over all values of $(x,x')$ such that $x\neq x'$. The interaction energy between two lines located at $x$ and $x+n$ is given by

\begin{widetext}
\begin{equation}
E_{int}^d(x,x+n) =
    M^z(x)\, M^z(x+n)  f_1(n) +    M^y(x)\, M^y(x+n) \left( f_1(n) - 3\, f_2(n) \right) \label{Eint2}
\end{equation}
\end{widetext}

\noindent where

\begin{equation}\label{f1}
    f_1(n) = \frac{1}{2L} \sum_{y,y'} \frac{1}{\left[n^2+(y-y')^2 \right]^{3/2}}
\end{equation}

\begin{equation}\label{f2}
    f_2(n) = \frac{1}{2L} \sum_{y,y'}\frac{(y-y')^2}{\left[n^2+(y-y')^2 \right]^{5/2}}
\end{equation}

\noindent In the limit $L \to\infty$ the sums in Eqs.(\ref{f1}) and
(\ref{f2}) can be evaluated using a continuum
aproximation\cite{YaGy1988} giving $f_1(n) \sim 1/n^2$ and $f_1(n) -3
f_2(n)\sim 0$. For $n>1$ the error in this approximation is smaller
than $0.1\%$. For $n=1$ they can be evaluated numerically giving
$f_1(1)=1.01243\ldots$ and $f_1(1) -3 f_2(1)=0.07276\ldots$. Then, Eq.(\ref{Eint2}) can be written as

\begin{eqnarray}
   E_{dip}^{int}(x,x+1)
   & =& M^z(x)\, M^z(x+1) + c_1\, \vec{M}(x).\vec{M}(x+1)  \nonumber\\
    &+& (c_2-c_1) M^y(x)\, M^y(x+1)\label{Eint-6}.
\end{eqnarray}

\noindent where $c_1=f_1(1)-1$ and $c_2=f_1(1) -3 f_2(1)$. Finally

 \begin{eqnarray}
e_{dip}^{int} &=& \frac{1}{L}\sum_{x\neq x'} \frac{M^z(x)\, M^z(x')}{|x-x'|^2} \nonumber \\
             &+& 2c_1\frac{1}{L}\sum_{x} \vec{M}(x).\vec{M}(x+1) +
    C\left[M^y(x)\right]
\label{Es4}
\end{eqnarray}

\noindent where $C\left[M^y(x)\right]$ is given by Eq.(\ref{correction}). Putting all these terms together we get Eq.(\ref{energia1}).

\subsection{Variational equations for a striped magnetization profile with sinusoidal wall (SWP) in the continuum limit}

\label{YGcontinuum}

In the continuum limit $h\gg 1$ and $w\gg1$ ($\delta\gg 1$) Eq.(\ref{energiabm}) can be written as

\begin{widetext}
\begin{equation}
e\left[\vec{M},\delta,\eta \right]= \delta - \delta \frac{1}{L} \sum_x cos\left( \phi(x)-\phi(x+1)\right) + M_0^2 \sum_{m=1,3,\ldots} b_m^2  D_m(h) - \frac{\kappa\, M_0^2}{2} \sum_{m=1,3,\ldots} b_m^2
\label{energiabm -cont}
\end{equation}
\end{widetext}

\noindent where $\kappa= \eta-3\, g+2(c_2-c_1)$  and the functions $D_m(h)$ are given in Eq.(\ref{Dm}). From Eqs.(\ref{perfilYG}) the Fourier cofficients $b_m$ in this limit are given by\cite{YaGy1988}

\begin{equation}\label{bm-YG}
    b_m = (-1)^{(m-1)/2} \frac{4}{\pi m}\frac{1}{1-m^2 \Delta^2}\, cos\left( \frac{\pi m \Delta}{2}\right)
\end{equation}

\noindent and\cite{YaGy1988}

\begin{equation}\label{sumbmYG}
    \sum_{m=1,3,\ldots} b_m^2 = (2-\Delta)
\end{equation}

\noindent with $\Delta=w/h$. The dipolar energy term in Eq.(\ref{energiabm -cont}) can be approached by\cite{YaGy1988}

\begin{equation}\label{edipYG}
    \sum_{m=1,3,\ldots} b_m^2  D_m(h)=  \left[\frac{\pi^2}{6}\,(2-\Delta) - \frac{\pi\, k}{2}\, G(\Delta)\right]
\end{equation}

\noindent where we have used Eq.(\ref{sumbmYG}), the cuadratic term in Eq.(\ref{Dm}) has been neglected, $k\equiv \pi/h$ and

\begin{equation}\label{GDelta}
    G(\Delta)\equiv \sum_{m=1,3,\ldots} m\, b_m^2(\Delta)
\end{equation}

\noindent Assuming a smooth profile inside the walls $|\phi(x)-\phi(x+1)| \ll 1$ the exchange term in Eq.(\ref{energiabm -cont}) can be approached by

\begin{equation}\label{eexchYG}
    e_{exc} \approx -\delta + \frac{\delta\Delta}{2w} \sum_{x=1}^w \left( \frac{d\phi(x)}{dx}\right)^2
\end{equation}

\noindent Taking $cos(\phi(x)) = cos(\theta)\, cos(\pi x/w)$ in the region inside a wall and replacing the summation in Eq.(\ref{eexchYG}) by an integral we have that

\begin{equation}\label{eexchYG2}
    e_{exc}= -\delta + \frac{\delta\, k^2}{2\Delta} (1- sin\, \theta)
\end{equation}

Replacing Eqs.(\ref{sumbmYG}), (\ref{edipYG}) and (\ref{eexchYG2}) into Eq.(\ref{energiabm -cont}) we get

\begin{widetext}
\begin{equation}
e_{SWP}= \frac{\delta}{2\Delta} k^2 (1-sin\,\theta) + \gamma \left(1-\frac{\Delta}{2} \right) cos^2\theta- \frac{\pi\, k}{2}\, G(\Delta)\, cos^2\theta
\label{energiaYG}
\end{equation}
\end{widetext}

\noindent where  $\gamma = \pi^2/3-\kappa$.
Minimizing Eq.(\ref{energiaYG}) respect to the variational parameters $(\theta,\Delta,k)$ we get

\begin{eqnarray}
   \frac{\delta\, k}{\Delta}\,(1- sin\, \theta) &=& \frac{\pi}{2}\, G(\Delta)\, cos^2\, \theta \label{YGEq1}\\
  \frac{\delta\, k^2}{\Delta^2}\,(1- sin\, \theta) &=& - \left[ \gamma + \pi\, k\, \frac{dG}{d\Delta}\right] \, cos^2\, \theta \label{YGEq2}
\end{eqnarray}

\begin{equation}
  \frac{\delta\, k^2}{2\Delta}\, cos\, \theta =  - \left[ 2\gamma
    \left(1-\frac{\Delta}{2} \right)- \pi\, k\, G(\Delta)\right] sin\,
  \theta\, cos\, \theta
\label{YGEq3}
\end{equation}

Notice that The fully saturated state $\theta=0$ is never a solution of the
  above equations, except in the limit $k\to 0$ (or
  $\Delta\to 0$), which corresponds to $\delta\to\infty$. On the other hand, the planar ferromagnetic state $\theta=\pi/2$ ($e=0$), is always solution of the above equations. For $\theta\neq \pi/2$, the variational equations reduce to Eqs(\ref{YGEq2-2}).

Close to the SRT (i.e., the transition between a state with $\theta=\pi/2$ and one with $\theta\neq\pi/2$) we can assume $cos\, \theta =s \ll 1$ and therefore $1- sin\, \theta = 1-\sqrt{1-s^2}\sim s^2/2$. Replacing into Eqs.(\ref{YGEq1})-(\ref{YGEq3})  they become

\begin{eqnarray}
    k &=& \frac{\pi \Delta}{\delta}\, G(\Delta) \label{YGEq1'} \\
  \frac{\delta\, k^2}{2\Delta^2} &=& - \left[ \gamma + \pi\, k\, \frac{dG}{d\Delta}\right] \label{YGEq2'}
\end{eqnarray}

\noindent independent of $s$, while Eq.(\ref{YGEq3}) becomes identically zero in the limit $s\to 0$ (SRT). Replacing Eq.(\ref{YGEq1'}) into Eq.(\ref{YGEq2'}) we find

\begin{equation}\label{YGEq4}
    \frac{\kappa}{\pi^2}= \frac{1}{3} + \frac{G(\Delta)}{2\delta} \left[G(\Delta)+2\Delta\, \frac{dG}{d\Delta} \right]
\end{equation}

Both $G(\Delta)$ and the expression between square brackets in Eq.(\ref{YGEq4}) are monotonously decreasing functions of\cite{YaGy1988} $\Delta$. Since the maximum allowed value is $\Delta=1$ (which corresponds to $w=h$. i.e., pure sinusoidal profile), the minimum value of $\kappa=\kappa_{min}$ for which a domain solution exists corresponds to $\Delta=1$. Using that\cite{YaGy1988} $G(1)=1$ and $(dG/d\Delta)_{\Delta=1}=-1$ we have

\begin{equation}\label{kappamin}
    \kappa_{min}=\pi^2\, \left( \frac{1}{3}-\frac{1}{2\delta}\right)
\end{equation}

From Eq.(\ref{YGEq1'}) this corresponds to $k=\pi/\delta$ or $h=\delta$. Replacing these values into Eq.(\ref{YGEq3'}) we see that in the limit $\kappa\to \kappa_{min}$ we have that $sin\,\theta \to 1$. Also from Eq.(\ref{energiaYG}) we see that in this limit the planar ferromagnetic and the domain solutions become degenerated. Hence, this point corresponds to the SRT and the SRT line in the $(\eta,\delta)$ space is given by Eq.(\ref{etaSRT}).

\section{Exact energy evaluation for a  striped magnetization profile with sinusoidal wall  (SWP) in a  lattice}
\label{YGdiscrete}

It is easier to carry out this calculation by considering a profile whose wall starts at $x=0$, i. e.

\begin{equation}
    M(x)=\left\{\begin{array}{ll}
             M_0 \, cos\left( \frac{\pi x}{w}\right) & if \;\; 0 \leq x \leq w \\
             -M_0 & if \;\; w\leq x \leq h
           \end{array} \right.\label{perfilYGDiscret}
\end{equation}

 \noindent where $h=1,2,\ldots$ and $w=1,2,\ldots,h$. The profile Eq.(\ref{perfilYG}) is related to the previous one by  $M^z(x)= M^z\left(x-(h-w)/2\right)$.  The profile  (\ref{perfilYGDiscret}) can be expanded as

 \begin{equation}\label{M(x)}
      M(x) = \frac{M_0}{2h} \, \sum_{m=1,3,\ldots}^{2h+1} c_m e^{i \frac{\pi m}{h} x}.
 \end{equation}

 \noindent so the coefficients $b_m$ of the expansion (\ref{Fourier-stripes}) are given by

 \begin{equation}\label{bmcm}
    b_m= \frac{1}{2h}{\cal R} \left[c_m\, e^{i\pi m\,(h-w)/2h} \right]
 \end{equation}

 \noindent where ${\cal R}[z]$ stands for the real part of $z$. The coefficients $c_m$ are given by

\begin{widetext}
 \begin{equation}
c_m = \frac{1}{M_0} \, \sum_{x=1}^{2h} \, M^z(x) \,
e^{-i \frac{\pi m}{h} x}
     = 2\, \sum_{x=0}^{w-1} \, cos\left( \frac{\pi x}{w} \right) \,
        e^{-i\frac{\pi m}{h} x}
        - 2\,  \sum_{x=0}^{h-1-w} \, e^{-i\frac{\pi m}{h} (x+w)}
\label{cm}
\end{equation}
\end{widetext}

 The summations in Eq.(\ref{cm}) can be carry out explicitly obtaining

 \begin{equation}\label{cm2}
    c_m = f_m^+ + f_m^- + f_m^0
 \end{equation}

\noindent where

\begin{equation}
f_m^{\pm} \equiv \sum_{x=0}^{w-1} \,  \left(e^{i \pi \alpha_{\pm}}\right)^x = \left\{
\begin{array}{ll}
  \frac{1-e^{i \pi \alpha_{\pm} w}}{1-e^{i \pi \alpha_{\pm}}} & \mbox{if $e^{i
      \pi \alpha_{\pm}} \ne 1$} \\
w & \mbox{if $e^{i \pi \alpha_{\pm}} = 1$}
 \end{array} \right.
\end{equation}

\noindent with $\alpha_{\pm}=\left(\pm \frac{1}{w}-\frac{m}{h} \right)$ and

\begin{eqnarray}
f_m^0 &\equiv& - 2\,  \sum_{x=0}^{h-1-w} \, e^{-i\frac{\pi m}{h} (x+w)}\nonumber\\
&=& 2\,  \frac{ e^{i\pi m w/h}+1}{e^{-i\pi m /h}-1}
\end{eqnarray}

The dipolar energy can then be evaluated from Eqs.(\ref{eDipInf}) and (\ref{Dm}). The anisotropy energy can be easily calculated and gives

\begin{equation}\label{aniYG-2}
   e_{an}=
   \left\{ \begin{array}{ll}
     - \kappa\, M_0^2   & \mbox{if $w = 1$}  \\
     - \kappa\, M_0^2 \left(1-\frac{\Delta}{2} \right) & otherwise
   \end{array}  \right.
\end{equation}

The exchange and correction terms in Eq.(\ref{energiabm}) can be expressed as

\begin{equation}
    e_{exc}=-\delta ' \frac{1}{h} \, \sum_{x=0}^{w-1} \vec{M}(x). \vec{M}(x+1)
  -\delta ' ( 1-\Delta)\, M_0^2
\end{equation}

\begin{eqnarray}
    C\left[M^y(x)\right]&=&  \frac{2(c_2-c_1)}{h} \, \sum_{x=0}^{w-1} M^y(x) M^y(x+1) \nonumber \\
    &+& 2(c_2-c_1) ( 1-\Delta)\, (1-M_0^2)
\end{eqnarray}

\noindent The summations in the above equations involve a finite number of terms that can be computed explicitly.

\section{Zero Temperature Monte Carlo Technique for striped domain patterns}
\label{MCmethod}

In order to check the different striped profiles used to minimize the
energy of the system, we implemented a simulated annealing protocol, based on Metropolis dynamics. The temperature was decreased down to zero at a constant rate $T(t)= T_0-r\,t$, where the time is measured in Monte Carlo Steps. All the simulations are made starting from a planar ferromagnetic state with $T_0=1$ and $r=10^{-4}$. Every simulation is repeated $100$ times using a different sequence of random numbers, to check for possible trapping in local minima. Since we are considering only periodic straight domains with Bloch walls, the problem is basically one dimensional and we can restrict the search to a one dimensional pattern over the $x$ direction fixing $M^x(x)=0$ and imposing periodic boundary conditions (PBC) in the $y$ direction. We also use PBC in the $x$ direction. In other words, we simulated a a lattice with $L_x \times L_y$ with $L_y=1$ and PBC, which are implemented by means of the Ewald sums technique.  For every set of values of $(\delta,\eta)$ we check the results for different  values of $L_x$ in order to avoid artificial frustration. We also performed some comparisons with MC results in a square $Lx=Ly$ lattice; the results were indistinguishable. This ansatz allows us to obtain MC results for values of $\delta$ up to $\delta=8$ (for which the maximum equilibrium value is $h=48$).

\section{Fourier coefficients for the Hyperbolic Tangent wall Profile (HWP)}
\label{Fouriertangh}

The function $\tanh(x)$ is very well approximated by
\begin{equation}
    \tanh(x)=\left\{\begin{array}{ll}
             x(1-\frac{x^2}{3}) & if \;\; 0 \leq x \leq \frac{1}{2} \\
             (1-e^{-2x})^2(1+e^{-4x}) & if \;\; \frac{1}{2} \leq x
           \end{array} \right.\label{perfilTanh}
\end{equation}

\noindent  Then the Fourier coefficients $b_m$ for the profile Eq.(\ref{hyperbolicprofile}) can be expressed as $b_m=b_m^1+b_m^2$, where

\begin{equation}\label{htanhdip}
    b_m^1 = \frac{4}{h_r} \int^{1/2}_{0} x(1-\frac{x^3}{3})\sin\left(\frac{m \pi x}{h_r}\right)dx
\end{equation}

\noindent and

\begin{equation}\label{htanhdip}
    b_m^2 = \frac{4}{h_r} \int_{1/2}^{h_r/2}  (1-e^{-2x})^2(1+e^{-4x}) \sin\left(\frac{m \pi x}{h_r}\right)dx
\end{equation}

\noindent where  $h_r\equiv h/l_w$. Both integrals can be solved analytically, leading to rather long expressions that can be handled with symbolic manipulation  programs.

\end{document}